\documentclass[conference]{IEEEtran}
\ifCLASSINFOpdf
  \usepackage[pdftex]{graphicx}
  \graphicspath{{plots/pdf/},{plots/schematics/}}
\else
\fi
\usepackage[cmex10]{amsmath}
\ifCLASSOPTIONcompsoc
  \usepackage[caption=false,font=normalsize,labelfont=sf,textfont=sf]{subfig}
\else
  \usepackage[caption=false,font=footnotesize]{subfig}
\fi
\begin{document}
	\newcommand{\argmax}{\operatornamewithlimits{argmax}}
\title{Deep Learning Based Transmitter Identification using Power Amplifier Nonlinearity}
\author{\IEEEauthorblockN{Samer S. Hanna and Danijela Cabric}
	\IEEEauthorblockA{Electrical and Computer Engineering Department,
		University of California, Los Angeles\\
		Email:  samerhanna@ucla.edu, danijela@ee.ucla.edu}
}

\maketitle

\begin{abstract}
The imperfections in  the RF frontend of different transmitters can be used to distinguish them. This process is called transmitter identification using RF fingerprints. The nonlinearity in the power amplifier of the RF frontend is a significant cause of the discrepancy in  RF fingerprints, which  enables transmitter identification. In this work, we use deep learning to identify different transmitters using their nonlinear characteristics.
By developing a nonlinear model generator based on extensive measurements, we were able to extend the evaluation of transmitter identification to include a larger number of transmitters beyond what exists in the literature. We were also able to study the impact of transmitter variability on identification accuracy. Additionally, many other factors were considered including modulation type, length of data used for identification, and type of data being transmitted whether identical or random under a realistic channel model. Simulation results were compared with experiments which confirmed similar trends. 
\end{abstract}


\IEEEpeerreviewmaketitle

\section{Introduction}

With the increase in the number of connected wireless devices, transmitter identification has become an important tool to stop malicious transmitters from impersonation. RF fingerprinting is a method that recognizes different transmitters using the device level differences in their RF frontends.  The circuit components of these frontends suffer from inherent variability due to the manufacturing process. Practically, two devices having the same specifications and even from the same manufacturer are slightly different in their RF characteristics. Unlike MAC address or other network layer identification protocols, these circuits nonidealities are device dependent and can not easily be forged. The imperfections that enable this differentiation between transmitters arise from clock jitter, digital to analog converters,
sampling errors, mixers or local frequency synthesizers, power amplifiers' non-linearity, device antennas, etc. The power amplifier's non-linearity is considered as the most significant source of differences~\cite{Wang_model_2015}. %

The idea of identifying RF transmitters based on their transmitted signals is not new. It started in the nineties with most of the work focused on using the transient behavior~\cite{trad_Danev_2012}.  The main drawback of these approaches is the need to locate the short transient segment of the transmitted signals. Later on, steady-state methods for RF identification were developed. These approaches eliminate the need for locating the transient part of the signal and include several features like modulation errors. %
More recently deep learning techniques have been proposed for RF identification. In these techniques, minimal processing is performed on the captured data and the machine learning technique would learn the features and classify the signal. In~\cite{ML_Kennedy_2008}, the frequency characteristics of signals are used as features, which are fed to a K-Nearest Neighbour classifier. The authors evaluated the effect of changing the SNR and the number of frequency bins. Fixed preambles collected from RF captures were used for recognition.
 A more comprehensive evaluation of the performance of RF identification was performed in~\cite{ML_frontend_Rehman_2012} and~\cite{ML_effect_impair_Rehman__2015}. In both these papers, the classification was performed on the frequency domain representation of the signal using neural networks. In the first, the ability of low and high-end receivers in distinguishing between different transmitter fingerprints was compared. In~\cite{ML_effect_impair_Rehman__2015}, the effect of training and testing under different channels was studied. Both papers used signal preambles collected from hardware captures for their work.
 In~\cite{ML_Youssef_2017}, an emphasis was placed on the machine learning technique that gives the highest identification accuracy. Classification methods like support vector machines, deep neural networks, convolutional networks, and a technique called multistage training were evaluated. Unlike previous work, the authors of~\cite{ML_Youssef_2017} relied on time domain features for recognition, although they showed that time-frequency features gave higher performance.  For comparison, they used random data obtained from RF captures. All the machine learning based approaches discussed in these papers, due to reliance on hardware captures, were not able to use more than 12 transmitters in their study.
 
In this paper, we compare several architectures of deep neural networks using frequency domain features. Unlike previous work, we developed a power amplifier nonlinearity model generator, which can generate as many realistic transmitter models as we want. This generator enabled us to explore RF identification using up to 500 transmitters and to evaluate factors like the similarity between transmitters, which are not possible using hardware.  We also extended our analysis to study the  impact of several parameters that were not considered in the previous work. A comparison between using similar data and random data for transmitter identification is performed. Additionally, we consider the effect of the length of the captured data on the recognition accuracy among other parameters.     

The rest of the paper is organized as follows. Section~\ref{sec:prob_desc} formally describes the RF identification problem. The proposed nonlinear model generator is discussed in Section~\ref{sec:gen_nonlin}, while in Section~\ref{sec:class}, we proposed a set of features for recognition and several classifiers. In Section~\ref{sec:sim}, simulations are performed to evaluate the aforementioned parameters. Experimental evaluation is used to corroborate our simulations in Section~\ref{sec:exp_eval}.
Section~\ref{sec:conc} concludes the paper.

\section{Problem Description}
\label{sec:prob_desc}
We assume there are $N$ transmitters, one of them is transmitting the complex digitally modulated signal $x(t)$. When transmitter $T_i$, where $i\in\{1,\cdots,N\}$, transmits this complex baseband signal, the RF output signal after its power amplifier is $f_i(x(t))$, where $f_i(x)$ models the nonlinear behavior of  RF frontend of transmitter $i$. Here we assume that the nonlinear function is transmitter specific, which is true in practice even if the transmitters have the same specifications and are from the same manufacturer. This signal is then sent through a channel and  the received signal is
\begin{equation}
	y(t) = h(f_i(x(t))) + n(t)
\end{equation}
 where $n(t)$ is a Gaussian random signal representing the added noise and $h(x)$ models the wireless channel. In this work, we consider an AWGN channel as well as a more realistic channel with timing errors, frequency errors, and fading. At the receiver side, $y(t)$ is sampled to obtain $y[n]$. The problem can be described as correctly identifying the transmitter given the received samples $y[n]$, which can be formulated as finding $\hat{i}$ defined as follows
 \begin{equation}
	 \hat{i} =	\argmax_j{ P(T_j=T_i | y[n])}
 \end{equation}  
under the assumption that the signal was sent by transmitter~$i$.  Here, we assume that the receiver does not have any prior information about the channel, the structure of the signal (if there is any structure), and the type of modulation.

For the purpose of getting the samples for transmitter identification that correspond to realistic radio frontends, there are two approaches: i) by capturing signal transmitted from hardware, or ii) by generating samples using simulation. Each of the two approaches has its advantages and its disadvantages. While real-world transmissions are similar to situations where transmitter identification systems are to be deployed, they impose some limitations on the analysis. First, practically the number of transmitters available for testing is limited by the hardware availability and its cost.  Second, the transmitters have a predefined set of nonlinearities that cannot be modified for experimentation purposes. Additionally, generating large datasets using real-world transmissions covering different parameters requires more effort than simulation. Simulations, on the other hand, give a larger flexibility and ease in testing, but the applicability of the results obtained is dependent on the accuracy of the model. In this paper, we use both simulations and hardware measurements to evaluate our transmitter identification algorithm.

To have the flexibility to explore the effect of the number of transmitters $N$ and the effect of modifying the variability between the nonlinear functions, we developed a nonlinear power amplifier model generator that realistically models the nonlinearity of real hardware. In particular, the statistical characteristics of the entire set of obtained functions should correspond to that of real hardware.

\section{Nonlinearity Model Generation}
\label{sec:gen_nonlin}

In this section, we describe our approach to building the nonlinear function generator used for the simulation. We start by selecting the nonlinear power amplifier model. In this work, we used the Saleh model \cite{saleh_frequency-independent_1981} which is defined by two parameters $\alpha$ and $\beta$ as described later.
To create such a generator, we measured the values of $\alpha$ and $\beta$  of many transmitters.  Then we fit the samples of these parameters to a statistical distribution, which will be later used to generate a family of new models with a tunable variability parameter.
 \begin{figure*}[!t]
 	\centering
 	\subfloat[]{\includegraphics{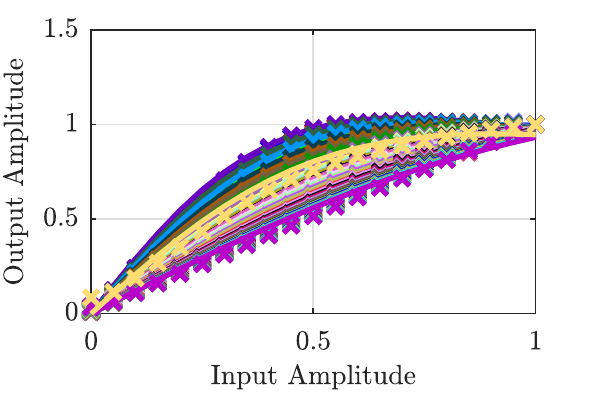}\label{fig:saleh_fit}}
 	\subfloat[]{\includegraphics{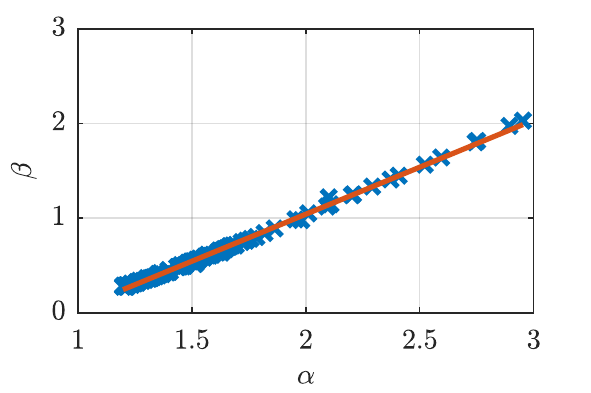}\label{fig:saleh_coef_dep}}
 	\subfloat[]{\includegraphics{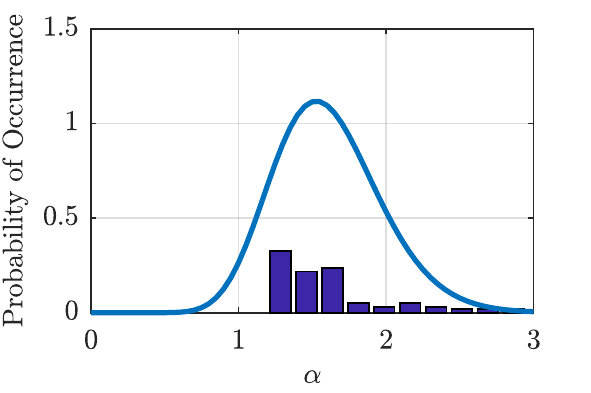}\label{fig:saleh_prob_dist}}
 	
 	\caption{Figure (a) shows the measured points as ticks, the solid lines represent the Saleh Model fits. Figure (b) shows in ticks the values of the $\alpha$ and $\beta$ parameters of the fitted models, while the solid line is the linear regression line. In Figure (c), the histogram of $\alpha$ and the fitted gamma distribution are shown. }
 \end{figure*}
 
\subsection{Measurements and Fitting}
 The nonlinearity measurements were performed using 8 USRPs. These USRPs are USRP2, USRP N200, and USRP N210 having different RF frontends namely: 2 SBX,  4 XCVR, 1 UBX, and 1 CBX . Throughout all the measurements, the USRP having the CBX front-end was used as a receiver, while the remaining USRPs were used as transmitters. To avoid external sources of  interference, the USRPs were connected using an SMA cable and a 40 dB attenuator. This value of attenuation along with the receiver gain were chosen to guarantee that we operate within the linear region of  the receiving USRP. Due to the limited number of USRPs and to increase the number of nonlinearity curves obtained, we ran the test using the same USRP at different frequency bands. This approach was inspired by the measurements in \cite{github_2017} showing that the nonlinearity varies with the center frequency. Using this fact,  101 nonlinearity curves were obtained from the 7 transmitters.

To obtain the AM-AM characteristic a two-tone test was used. The transmitter and the receiver were both tuned to a center frequency in the range of both RF frontends while having a 1 MHz bandwidth. The transmitter started generating two equal amplitude sine waves at  24 and 36 kHz while the receiver measured the received power. The amplitude was varied from 0 to 1 using a step of $0.05$. The obtained points after normalization are shown as crosses in Figure~\ref{fig:saleh_fit}.

 In the literature, there exist many functions used to capture the power amplifier nonlinearity among which are Ghorbany~\cite{ghorbani_effect_1991}, Rapp~\cite{rapp_effects_1991}, and Saleh~\cite{saleh_frequency-independent_1981} models, from which we found that the Saleh model is the most suitable for our purposes. The Saleh model relates the input amplitude $r$ with the output amplitude $A(r)$ using the following equation
\begin{equation}
A(r) = \alpha r / (1 + \beta r^2),
\end{equation}
where $\alpha$ and $\beta$ are the model parameters\footnote{We only considered the amplitude component of the nonlinearity.}. The fit was performed using the least squares equations from \cite{saleh_frequency-independent_1981}. The fitted curves are shown in solid lines in Figure~\ref{fig:saleh_fit}. Our selection for the Saleh model was based on a tradeoff between how well the model fits the data and the number of parameters. We looked for the model with the fewest number of parameters so that it would be easy to analyze their statistical distribution. Using this distribution, we generate new models to emulate different RF fingerprints.

\subsection{Model Generation}
After obtaining 101 pairs of $\alpha$ and $\beta$ values, we analyzed their statistics. The scatter plot of the pairs of values shown in Figure~\ref{fig:saleh_coef_dep} indicates a linear correlation between the two values. Using linear regression, we are able to relate the values of $\beta$ and $\alpha$. Next, we fit the values of $\alpha$ to a probability distribution.  The histogram of the values of $\alpha$ is shown in Figure~\ref{fig:saleh_prob_dist}. A gamma distribution, with mean $\mu$ and standard deviation $\sigma$, was fitted to the histogram which is also shown in Figure~\ref{fig:saleh_prob_dist}.  The tails of the gamma distribution less than 1 and more than 3 were removed to avoid unrealistic values. To simulate changes in the variability in the RF frontend, when generating new values of alpha and beta, we used $\sigma_s$ as standard deviation, which is defined as
$\sigma_s=s\sigma$, where we call $s$ as the coefficient of transmitter variability.
For example, for a low-cost device with high variability, the transmitters would have a larger coefficient of transmitter variability.

\begin{figure}[!tb]
	\subfloat[]{\includegraphics{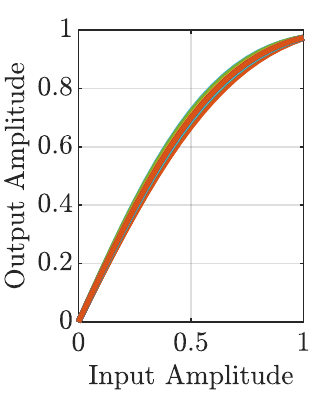}\label{fig:saleh_rand_10}}
	\subfloat[]{\includegraphics{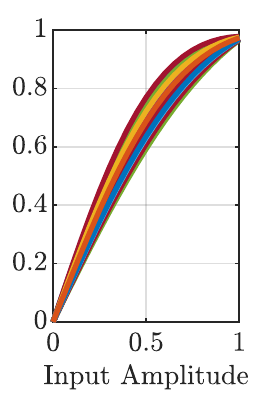}\label{fig:saleh_rand_100}}
	\subfloat[]{\includegraphics{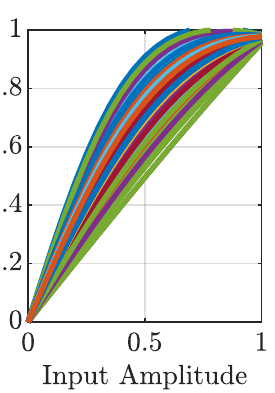}\label{fig:saleh_rand_1000}}
	\caption{Figures (a)-(c) show 100 examples of nonlinearity curves obtained from the generator for coefficients of variability: 0.01, 0.1, and 1 respectively.\label{fig:saleh_rand}}
\end{figure}

So the process of generating a model is as follow (1) Sample the gamma distribution with the calculated mean  and the scaled standard deviation to obtain $\alpha$, (2) Use the fitted line relating $\beta$ and $\alpha$ to calculate $\beta$. Figures~\ref{fig:saleh_rand_10}, ~\ref{fig:saleh_rand_100}, and \ref{fig:saleh_rand_1000} show examples of generated nonlinearity curves for different coefficients of transmitter variability taking values of 0.01, 0.1, and 1. 

\section{Classifier Features, Structure, and Parameters}
\label{sec:class}
In this work, we use the frequency representation of the received symbols as features for transmitter identification. The set of received samples is divided into windows of size 256. The discrete Fourier transform of each window is calculated and the obtained Fourier coefficients are averaged. The result of this operation is 256 complex coefficients. Several methods to feed this data to our classifier were tested: the magnitude only representation, the cartesian representation, and the polar representation. For the first, the input dimensions are $256\times1$ and for second and third the input dimensions are $256\times2$.

\begin{figure}[t!]
	\centering
	\subfloat[Fully Connected Net]{\includegraphics{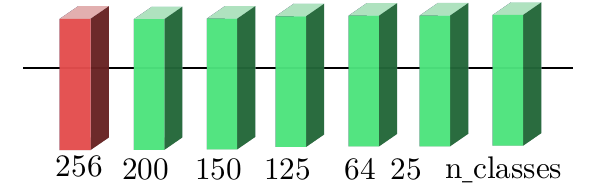} \label{fig:ft_net_0}}
	\\[-0.5ex]
	\subfloat[Convolutional Net]{\includegraphics{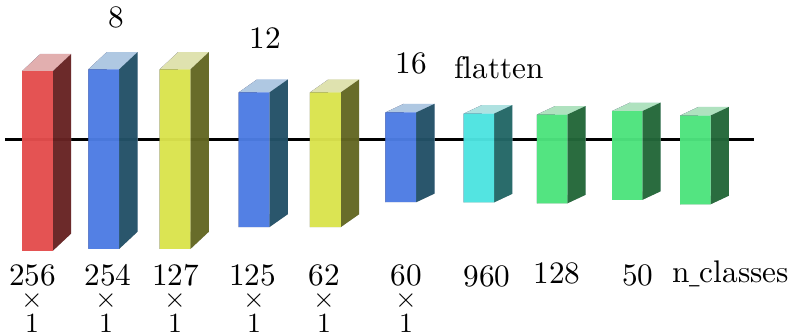} \label{fig:ft_net_2}}
	
	\caption{Architectures of neural networks that were evaluated when using the magnitude. The colors indicate the type of layer: red for input, green for dense, blue for convolutional, yellow for maxpool, and Cyan for flatten. The dimension of the output of each layer is written underneath. The number of filters is written above the convolutional layers.}
\end{figure}

As for the classifier, we used deep neural networks. Deep neural networks are widely used in computer vision and recently has been used to address several problems in communications \cite{ML_dl_physical_oshea_2017}.
  Since our feature size is small (256 or 512), we are able to use fully connected networks. Hence, the two architectures we evaluated are a fully connected network and a convolutional neural network. The architectures of both networks used for the 256 input are shown in Figures \ref{fig:ft_net_0} and \ref{fig:ft_net_2}. For the first architecture, several layers of fully connected neural networks were used with the number of neurons decreasing as we move from one layer to the next. For the second architecture, two 3$\times$3 convolutional networks followed by 2$\times$1 maxpooling were used. A dense layer was then used for classification. These architectures were modified for the 512 featues size as follows: for the fully connected network, one more layer of size 300 was added and for the convolutional network, the convolution was performed on two dimensions instead of one.

All the training and testing was done using the Keras API for Tensorflow~\cite{keras_2015}. A batch size of 32 was used. The loss function chosen is categorical cross entropy. For training, the Adam~\cite{adam_2014} optimizer was used with the default values provided in Keras. An epoch consisted of 1000 samples per class. A maximum of 25 epochs was used with early termination occurring if the loss did not decrease for 5 epochs. Several neural networks were trained and the one giving the best performance was kept.

\section{Simulations based Study}
\label{sec:sim}
Through simulations, we studied the performance of different neural network architectures and methods to represent the signals. After choosing the best performing architecture, the effect of two sets of parameters on classification accuracy was studied. The first set includes transmitter specific parameters, which include the number of transmitters and the variability between transmitters. The second set consists of different signals and channel parameters, which include SNR, channel model, modulation type, and packet length.

\subsection{Data Generation}
The main benefit of simulation is the infinite amount of data that can be generated. This benefit will be drastically hindered if we generated a limited amount of data and used it for training or testing. Hence, we resorted to online sample generation which works as follows. At first, we generate the required number of RF frontend nonlinearities, which is the number of required transmitters. Once a training sample is needed, data is generated with the specified packet length, modulated using the specified modulation type, then passed through one of the transmitter nonlinearities that constitute the label for the neural network. Afterwards, the channel impairments are modeled and at last  the chosen feature is calculated. The labeled samples are then fed to the neural network. All these steps are done online, i.e, the labeled samples are generated as needed and are not stored nor fed to the network more than once. This approach avoids the problem of overfitting. Overfitting occurs when the network learns features specific only to the training set. Since we are using each sample only once, this problem is entirely avoided.  

Unless stated otherwise, the following parameters were used for simulations. The generated data consisted of packets of 8192 symbols. Each symbol was modulated using QPSK having two samples per symbols and root raised cosine pulse shaping with 0.2 excess bandwidth. The default value of SNR was 20 dB and the default number of transmitters was 20. Two types of input data were investigated; the first is random data where each time a packet consists of completely random symbols. As for the second type, referred to as same data, the same random packet is used each time for training and testing.

Two channel models were investigated. The first one is the AWGN channel. The second one is a dynamic channel, which includes a set of more realistic impairments including timing errors, frequency errors, fading, and noise. The timing error is simulated by interpolating the signal by a factor of 32, choosing a random offset, and then downsampling. As for the frequency error, it is obtained by multiplying the signal with a complex exponential. The frequency of the exponential is selected from a Gaussian distribution with zero mean and a standard deviation of 1 kHz. For fading, a three tap channel was used along with a Rayleigh coefficient of scale 0.5. Finally, the noise is added to the signal.

 Note that, as stated earlier, data is generated online. So, each epoch consisted of different samples. Hence, the size of the training sets would vary between 5,000 to 25,000 times the number of classes depending on the loss value of each epoch (since we are using early termination as stated earlier).  As for the testing, 1000  samples per each class were generated and used to evaluate the performance. 

\subsection{Feature and Neural Network Architecture}
We start by comparing  methods to represent the data and  neural network architectures to classify the transmitters. Once, we determine the feature and architecture giving the best performance, we will use them for the rest of the paper. 

As stated earlier, the captured IQ samples were processed as chunks of length 8192 symbols unless otherwise stated. Several methods to feed these packets to the neural network are considered. All of them are based on the frequency representation of the data. An FFT transform of length 256 was chosen. The first representation is the magnitude of the FFT only, the second is the cartesian representation, and the  third is  the polar representation.

Each of these representations was used as an input to the two proposed neural network architectures. The first architecture is the fully connected neural network shown in Figure \ref{fig:ft_net_0}, while the second architecture is the convolutional network shown in Figure \ref{fig:ft_net_2}. 

\begin{figure}[t!]
	\centering
	\includegraphics{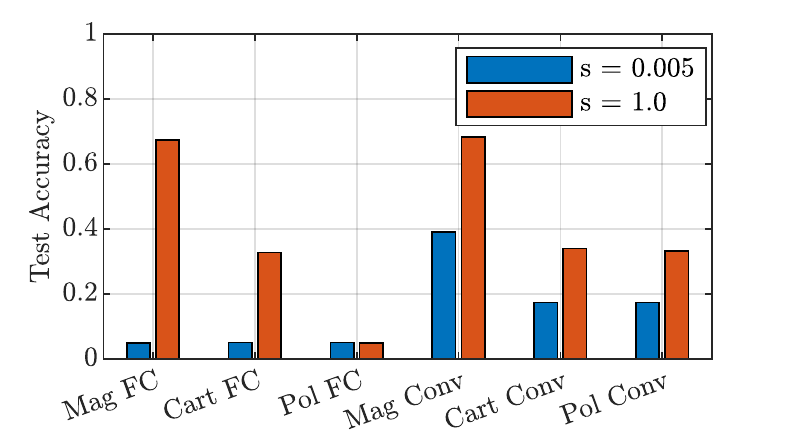}
	\caption{A comparison between the different inputs and architectures. At low variability only the magnitude input and convolutional architecture network is better than random guess. At larger variability, it still gives the best performance, slightly outperforming the fully connected network with magnitude input. $s$ is the coefficient of variability. }
	\label{fig:arch}
\end{figure}

The evaluation of these representations and these networks was done on data originating from two sets of transmitters, one has a coefficient of variability of 0.005 and the other of~1. Both results are shown in Figure \ref{fig:arch}. Results show that for transmitters that have similar nonlinearities using only the magnitude as a feature and the convolutional neural network gives the best recognition accuracy. As for transmitters that are more disparate, using the magnitude only gives the best performance, while the convolutional neural network is slightly better than the fully connected one. Based on these results, we decided to use the convolutional neural network shown in Figure \ref{fig:ft_net_2} with the magnitude of the Fourier transform as the input feature for the rest of the paper.

\subsection{Transmitter Parameters}
After determining the best performing architecture and feature, we study the effect of variability between transmitters and the number of the transmitters on the classification accuracy. 
\subsubsection{Coefficient of Transmitter Variability}
\begin{figure}[t!]
	\centering
	\includegraphics{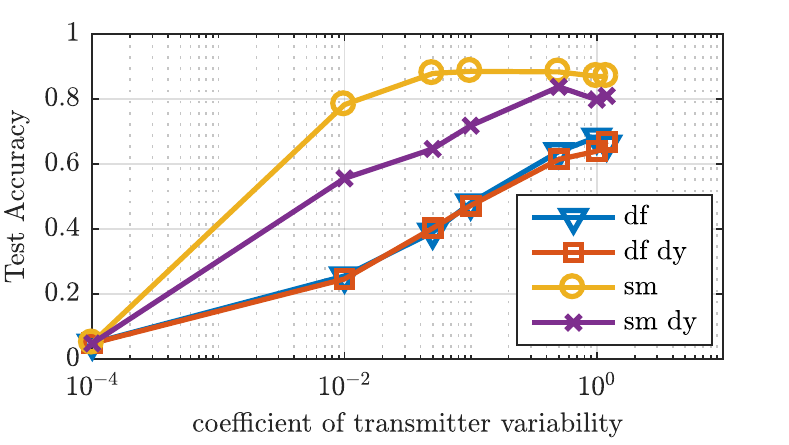}
	\caption{The effect of changing the spread between different transmitters obtained by changing the coefficient of variability of the generator. We can see that as the variability increases the ability of the network to identify RF fingerprints improves.}
	\label{fig:nl_scale}
\end{figure}

As discussed in Section \ref{sec:gen_nonlin}, our nonlinearity model generator enables us to control the simulated variability between models. An example of the effects of the coefficient of variability on the AM-AM curves was shown in Figure~\ref{fig:saleh_rand}. From Figure~\ref{fig:nl_scale}, we see that as the variability between transmitters decreases, the ability of the neural networks to distinguish between transmitters decreases. We found that the same packet inputs give a better identification accuracy. This improvement is because the neural network does not have to account for the difference in the data as when using random packets. From Figure~\ref{fig:nl_scale},  we can also see that the general trend is that performance improves as we increase the  variability between transmitters.    %
As for the effect of channel impairments, we can see that for lower values of variability and random data the impairments degrade the performance by up to 20\%. But, for the same packet and larger value of transmitter variability the difference becomes insignificant. This result shows that when the underlying data is random and the difference between transmitters is small, the neural network is highly affected by the additional channel impairments. In  Figure~\ref{fig:nl_scale} and all the subsequent figures, we refer to curves representing results for same packets by ``sm" and different packets by ``df". ``dy" refers to the dynamic channel else the AWGN channel was used.

\subsubsection{Number of Transmitters}
\begin{figure}[t!]
	\centering
	\includegraphics{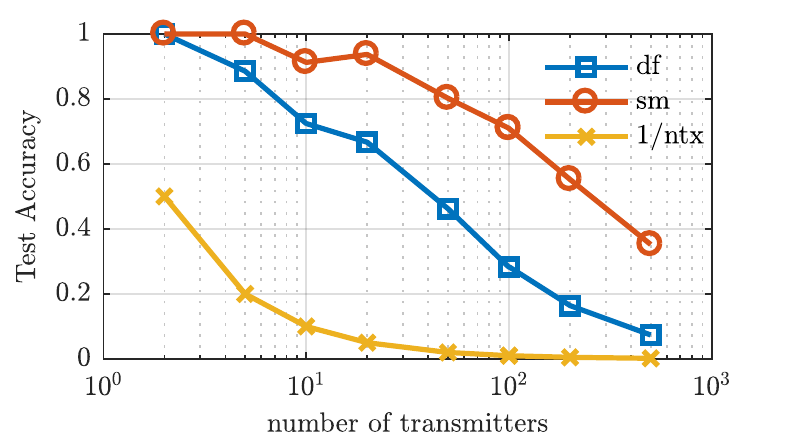}
	\caption{The effect of changing the number of transmitters. As the number of transmitters increases the accuracy decreases. Yet, it is still better than the random guess (yellow curve). Up to 500 hundred transmitters were evaluated, which would be hard to realize using real hardware.}
	\label{fig:ntx}
\end{figure}

Next, we analyze the performance of both types of data against the number of transmitters. From Figure \ref{fig:ntx}, we can see that as we increase the number of transmitters the classification accuracy drops. In Figure~\ref{fig:ntx}, we plotted a line showing the results obtained if we had used random guessing, which is equal to the inverse of the number of transmitters. So, although the recognition accuracy decreases with the number of transmitters, it is still higher than using random guessing. %

\subsection{Signal and Channel Parameters}
After investigating the effect of the number of transmitters and variability, we investigate the effects of SNR, channel model, the length of packets, and modulation type. 

\subsubsection{SNR}
\begin{figure}[t!]
	\centering
	\includegraphics{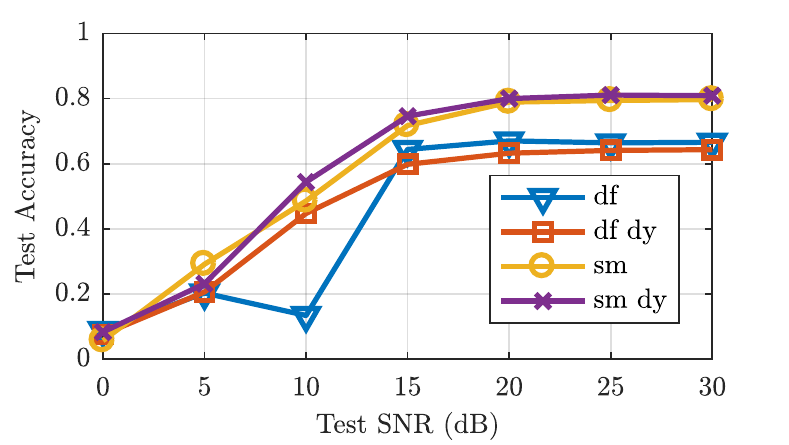}
	\caption{The effect of changing the SNR. As the SNR increases, the performance of our system improves.}
	\label{fig:mx_snr}
\end{figure}

The effect of the SNR on the recognition is first evaluated. In any practical scenario, the receiver has no control over the value of the SNR of the received signal. Hence, the most practical strategy is to train over signals having different values of SNRs. The training data were chosen to have SNR randomly selected from values from 0 dB to 30 dB with step~1. This training set should provide us with a robust neural network capable of operating over a wide range of SNRs. For the testing stage, we tested the network over SNRs from 0 dB to 30 dB having a step of 5dB. AWGN and dynamic channels were investigated, also using random data and same data packets.
The breakdown of the results for different SNRs is shown in Figure~\ref{fig:mx_snr}. 
From these figures, we can see that the performance when using the same packet is better than when using random data.  As for the comparison between AWGN and dynamic channels, we can see that the results are close. Thus, training the network using samples with multiple impairments does not have a major impact on its ability to differentiate between transmitters.

\subsubsection{Packet Length}
\begin{figure}[t!]
	\centering
	\includegraphics{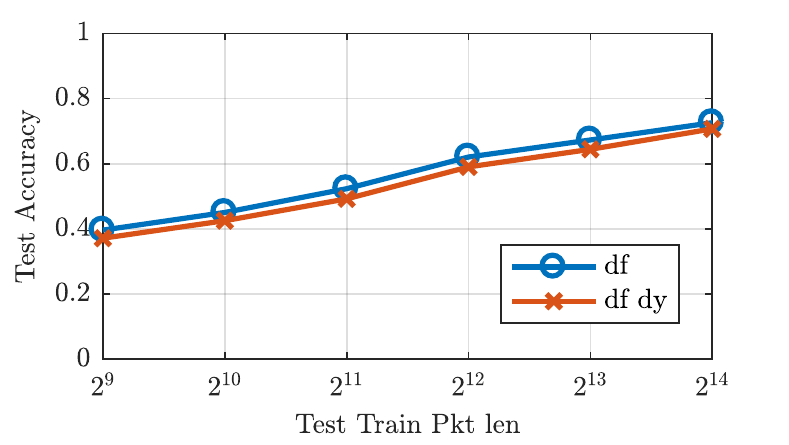}
	\caption{The effect of changing the packet length used to obtain the frequency representation. As the packet becomes longer,  the performance improves.}
	\label{fig:pkt_len}
\end{figure}

Then, we evaluate the effect of the packet length. As we have control of the length of the capture used in transmitter identification, the training and testing were done on packets of the same size. The results obtained are plotted in Figure~\ref{fig:pkt_len}. Each point represents the evaluation of a neural network trained with packets of that length. We can see that as we increase the length of the packet the recognition accuracy improves. The network performance in the noise only channel was slightly better than with the impairments.

\subsubsection{Modulation Type}
\begin{figure}[t!]
	\centering
	\includegraphics{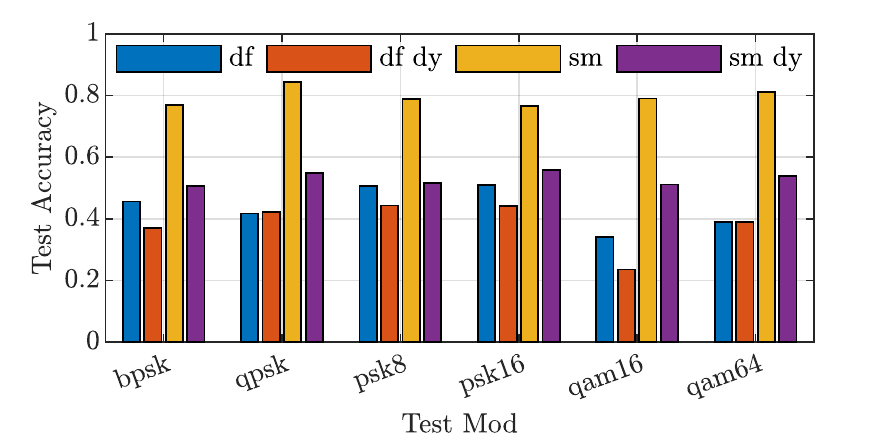}
	\caption{The effect of changing the modulation type. Our trained network was able to successfully classify different modulation types.}
	\label{fig:mod_mx}
\end{figure}

The last signal parameter we investigated is the type of modulation. Our approach was to train the neural network over many types of modulation. In particular, one network was trained over a mixture of all several types of modulations, while testing was done for one modulation at a time. 
 The breakdown of the results with respect to modulation type for different and same packets are shown in Figure ~\ref{fig:mod_mx}.
 We can see that the neural network is able to perform well for different modulation types.


 \section{Experimental Evaluation}
 \label{sec:exp_eval}
 So far the results were solely based on simulations. In this section, we evaluate the classifier using hardware captures. 
 \subsection{Measurement Procedure}
 The USRPs previously mentioned were used using the same setup discussed in Section~\ref{sec:gen_nonlin}. All USRPs were set to use only one frequency. Hence, we only used 7 transmitters. Each transmitter was set to send signals having different amplitudes. The results were captured by the receiver and stored. Two types of packets were used; the first is completely random and the second consists of the same random vector of length 1024 repeated over and over. A total of 40920 packets was used, 80\% for training and 20\% for testing.

\begin{figure}[t!]
 	\centering
 	\includegraphics{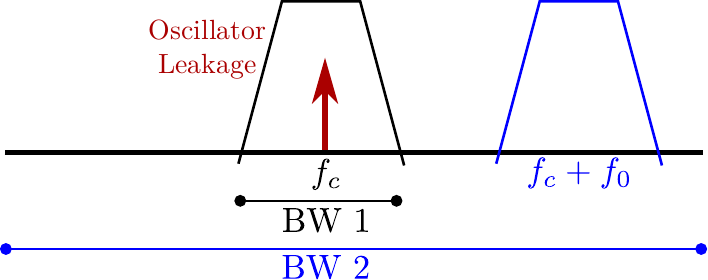}
 	\caption{The data capture setup. In the first, the oscillator leakage was included, while in the second (shown in blue) it was avoided.}
 	\label{fig:usrp_setup}
 \end{figure}

 Two signal captures were used. In the first, both the signal and the capture have  the same center frequency $f_c$ (2.4 GHz) and bandwidth "BW 1" (1 MHz). In this setup, the oscillator leakage from both the transmitter and receiver is included in the capture. We will refer to this setup as \textit{center}. In the second setup, the capture used a bandwidth "BW 2" (5 MHz) and the center frequency $f_c$, while the signal had a bandwidth of 1 MHz and a center $f_c+f_0$ ($f_0$ used is 1.25 MHz). At the receiver end, a frequency translating filter is used followed by decimation to bring the signal back to baseband. This setup is referred to as \textit{xlat}. The two setups are illustrated in Figure~\ref{fig:usrp_setup}.   

 \subsection{Results}
\begin{figure}[t!]
 	\centering
 	\includegraphics{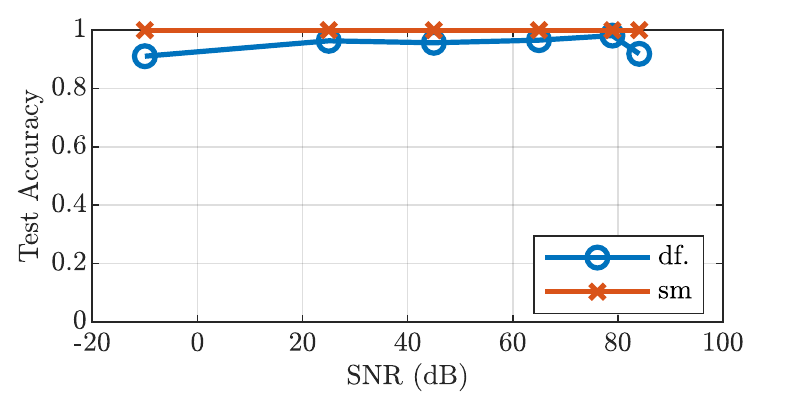}
 	\caption{The effect of changing the SNR when using the center setup. Due to oscillator leakage, recognition is not affected by signal amplitude. The -10 dB point represents no signal ($-\infty$ SNR) altered to fit on the plot.}
 	\label{fig:usrp_txamp_center}
 \end{figure}

\begin{figure}[t!]
 	\centering
 	\includegraphics{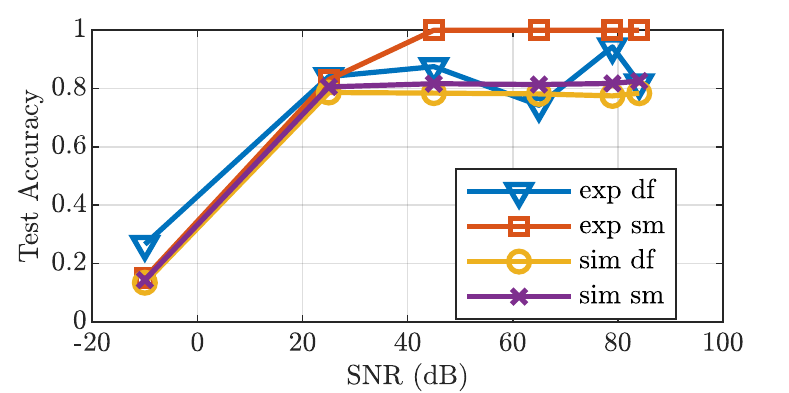}
 	\caption{The effect of changing the amplitude of the input when using the xlat setup. The -10 dB point represents no signal ($-\infty$ SNR) altered to fit on the plot. ``exp" stands for experimental results and ``sim" for simulation.}
 	\label{fig:usrp_txamp_xlat}
 \end{figure}

 For the experimental evaluation, we measured the effect of changing the SNR of the signal by changing the amplitude of the transmitted signal. The network was trained over all SNRs and the performance was tested for each SNR alone. The results are shown in Figures \ref{fig:usrp_txamp_center} and \ref{fig:usrp_txamp_xlat} for the center and xlat setups. The first point tested was zero amplitude (It was plotted as -10 dB for illustrative purposes), i.e. no signal is transmitted. Yet, we see in Figure~\ref{fig:usrp_txamp_center} that the recognition accuracy is close to 100\%.
  While, from Figure~\ref{fig:usrp_txamp_xlat}, we see that  we get an accuracy close to 1/7 for the same zero amplitude once we don't include oscillator leakage. This shows that our network is using the oscillator leakage to identify transmitters in the center setup. In Figure~\ref{fig:usrp_txamp_center}, we see that the center setup has an almost constant accuracy. While in Figure~\ref{fig:usrp_txamp_xlat}, the accuracy increases quickly and saturates. For comparison, a simulation using a dynamic channel with the same setup as in Section~\ref{sec:sim} was performed. For a fair comparison, the number of transmitters was set to seven and  SNR values were based on those used in our experiment. The results are plotted in Figure \ref{fig:usrp_txamp_center}. From this Figure, we can see that the experimental evaluation and simulations results are  quite similar for the random packets. However, the experimental results are better for the same packet. Note that there are slight differences between the input data in the simulations and the experiment. First, the packet structure for experimental repeats every 1024 and for the simulation there is no such repetition within the packet. Also, the simulation includes Rayleigh fading, while during the experiment the USRPs were connected using a wire. Overall, these results -- to some extent -- verify that the nonlinearity models obtained from our generator and the corresponding simulation results are comparable to the real world measurements. 


\section{Conclusion}
\label{sec:conc}
This paper provides a comprehensive study of transmitter identification based on RF nonlinearity fingerprints using deep learning. A nonlinear model generator was developed by statistically fitting multiple measured nonlinearity curves from USRPs. Nonlinearity models obtained from this generator were used to perform a comprehensive evaluation of the performance of RF fingerprinting. Results showed that a convolutional neural network with the magnitude of the frequency representation of the data gives the best performance. As the number of transmitters increases or the variability of nonlinearity decreases, the ability of our system to correctly identify different transmitters decreases. As for data format, same data packets outperform random data, while increasing the length of the used capture steadily improves performance. RF fingerprinting was shown to perform well under different modulation techniques. Simulations under AWGN channels gave better performance than dynamic channels. Experimental results confirm the trends observed via simulation, which shows the effectiveness of our model as well as the good performance of our classifier under realistic channel impairments. 
\section*{Acknowledgment}
This material is based upon work supported by the National Science Foundation under Grant No. 1527026.

\bibliographystyle{IEEEtran}
\bibliography{IEEEabrv,mybib}
\end{document}